\documentstyle[12pt]{article}

\newcommand{\journal}[4]{{\em #1~}#2\,(19#3)\,#4;}

\newcommand{\hpa}{\journal {Helv. Phys. Acta}}

\newcommand{\ijmp}{\journal {Int. J. Mod. Phys.}}

\newcommand{\cmp}{\journal {Comm. Math. Phys.}}

\newcommand{\np}{\journal {Nucl. Phys.}}
\newcommand{\pl}{\journal {Phys. Lett.}}

\newcommand{\prep}{\journal {Phys. Rep.}}

\newcommand{\annp}{\journal {Ann. Phys. (N.Y.)}}

\renewcommand{\a}{\alpha}
\renewcommand{\b}{\beta}
\newcommand{\g}{\gamma}           
\renewcommand{\d}{\delta}         \newcommand{\D}{\Delta}
\newcommand{\e}{\varepsilon}

\renewcommand{\l}{\lambda}        \newcommand{\LA}{\Lambda}
\newcommand{\m}{\mu}
\newcommand{\n}{\nu}
\renewcommand{\o}{\omega}         \renewcommand{\O}{\Omega}
             \newcommand{\Ps}{\Psi} 
\newcommand{\r}{\rho}
\newcommand{\s}{\sigma}           \renewcommand{\S}{\Sigma}
         \newcommand{\T}{\Theta}
\newcommand{\f}{{\phi}}           

\newcommand{\x}{\xi}              
       \newcommand{\Y}{{\Upsilon}}

\newcommand{\LL}{{\cal L}}
\newcommand{\MM}{{\cal M}}

\newcommand{\SS}{{\cal S}}

\newcommand{\VV}{{\cal V}}
\newcommand{\WW}{{\cal W}}

\newcommand{\Sla}{\raise.15ex\hbox{$/$}\kern -.70em D}

\newcommand{\lac}{\left\{}\newcommand{\rac}{\right\}}

\newcommand{\complex}{{\kern .1em {\raise .47ex
\hbox {$\scriptscriptstyle |$}}
    \kern -.4em {\rm C}}}
\newcommand{\real}{{{\rm I} \kern -.19em {\rm R}}}
\newcommand{\rational}{{\kern .1em {\raise .47ex
\hbox{$\scripscriptstyle |$}}
    \kern -.35em {\rm Q}}}
\renewcommand{\natural}{{\vrule height 1.6ex width
.05em depth 0ex \kern -.35em {\rm N}}}

\newcommand{\half}{\frac 1 2}
\newcommand{\pa}{\partial}

\newcommand{\sla}{\raise.15ex\hbox{$/$}\kern -.57em}
\newcommand{\twiddle}{\lower.9ex\rlap{$\kern -.1em\scriptstyle\sim$}}

\newcommand{\equ}[1]{(\ref{#1})}
\newcommand{\eq}{\begin{equation}}
\newcommand{\eqn}[1]{\label{#1}\end{equation}}
\newcommand{\eea}{\end{eqnarray}}
\newcommand{\eqa}{\begin{eqnarray}}
\newcommand{\eqan}[1]{\label{#1}\end{eqnarray}}
\newcommand{\ba}{\begin{array}}
\newcommand{\ea}{\end{array}}
\newcommand{\eqac}{\begin{equation}\begin{array}{rcl}}
\newcommand{\eqacn}[1]{\end{array}\label{#1}\end{equation}}

\renewcommand{\=}{&=&} 
\renewcommand{\in}{\int d^4x}
\newcommand{\ds}[1]{\frac{\delta\Sigma}{\delta {#1}}}

\newcommand{\dd}[1]{\frac{\delta}{\delta {#1}}}

\newcommand{\non}{\nonumber \\}
\newcommand{\6}{\partial}

\renewcommand{\title}[1]{\null\vspace{25mm}

\noindent{\Large{\bf #1}}\vspace{10mm}

\noindent { }}
\newcommand{\authors}[1]{\noindent {\large #1}\vspace{3mm}

}
\newcommand{\address}[1]{\noindent #1\vspace{5mm}

}
\renewcommand{\abstract}[1]{\vspace{10mm}

\noindent{\small{\em Abstract.} #1}\vspace{2mm}

} 

\begin{document}


\hspace*{\fill} REF. TUW 98-06
\begin{center}
\title{A Note on the $4D$ Antisymmetric Tensor \\
       Field Model in Curved Space-Time}
\authors{H. Zerrouki\footnote{Work supported in part by the
         ``Fonds zur F\"orderung der Wissenschaftlichen Forschung''
         under Contract Grant Number P11582-PHY.}}
\address{Institut f\"ur Theoretische Physik, 
         Technische Universit\"at Wien\\
         Wiedner Hauptstra\ss e 8-10, A-1040 Wien (Austria)}         
\end{center}

\abstract{ Using the BRS techniques, we prove the existence of a 
local and nonlinear symmetry of the gauge fixed action of the 
antisymmetric tensor field model in curved background. 

}


\setcounter{page}{0}
\thispagestyle{empty}


\newpage
\section{Introduction}

Topological field theories \cite{birm1} are mathematically as well
as physically interesting \cite{wit}. In four 
dimensions we have two interesting topological field theories, The 
topological Yang--Mills model \cite{wit} and the antisymmetric tensor 
field model \cite{hor1}. The algebraic renormalization\footnote{More 
details about the subject of algebraic renormalization can be found in 
the last reference of \cite{pig3}} of the topological Yang--Mills field
theory was carried out in \cite{mwo} and then extended to curved
space--time in \cite{zer2}. On the other hand, the algebraic 
renormalization of the antisymmetric tensor field model (in the flat
space--time limit) was first done in \cite{sor2} and then generalized 
to a curved space--time admitting a covariantly constant vector field 
in \cite{zer3}.
In this work we will go a step further and try to generalize the 
analysis of \cite{zer3} to an arbitrary, curved, Riemannian manifold.\\  
The paper is organized as follows, in section 2 we give the 
gauge fixed action and display the BRS transformations of 
all the fields appearing in the model. Next, in section 3 
we derive the on--shell local supersymmetry--like transformations, i.e.
the anticommutator of the BRS operator and of the local
susy--like operator leads to Lie derivative . In 
section 4 we extend the on--shell analysis (of section 3) to
the off--shell level. Here we will see that the local 
susy--like Ward operator, when it acts on the total action describing
the model, gives rise to a hard breaking which is quadratic in the 
quantum fields. In order to eliminate this quadratic breaking we introduce 
auxiliary fields. The result of this construction is that the 
anticommutator of the linearized Slavnov operator
and the Ward operator of the new symmetry does not close on 
diffeomorphisms for
certain fields. 
Furthermore, the most important fact is that the 
total action is invariant under this
new, local, nonlinear (and not susy--like) symmetry. \\
On the other hand,
one could use this symmetry (first one has to show its
validity at all order of perturbation theory) to
prove the finiteness of the $4D$ antisymmetric tensor field model in 
a general class of curved manifolds: generalizing the results of 
\cite{zer3}. 

\section{The model}  \label{sec1}

First let us consider the following classical action in curved 
space--time
\eq
\S_{inv}= - \frac{1}{4}\int_\MM d^4x ~\e^{\mu\nu\rho\sigma} 
F^a_{\mu\nu}B^a_{\rho\sigma} \ ,
\eqn{curv-bf}
where $\MM$ is a curved manifold described by the Euclidean metric
$g_{\m\n}$. 
The field strength is described by
\eq
F_{\m\n}^a = \pa_\m A^a_\n - \pa_\n A^a_\m + f^{abc} A^b_\m
             A^c_\n,
\eqn{strengthfga}
where $A^a_\m$ is the gauge field which belongs to the adjoint 
representation of a compact Lie group whose structure constants
are denoted by $f^{abc}$. The antisymmetric tensor field $B^a_{\m\n}$
is also Lie algebra valued and $\e^{\m\n\r\s}$ is the Levi--Civita
tensor density\footnote{In this paper we denote the inverse 
of the metric by $g^{\m\n}$ and its determinant by $g$. 
Under diffeomorphisms, $\sqrt g$ behaves like a scalar density
of weight +1, whereas 
the volume element density $d^4x$ has weight -1.
The Levi--Civita antisymmetric tensor density 
$\e^{\m\n\r\s}$ has weight $+1$ and $\e_{\m\n\r\s}$ has weight $-1$.
Furthermore, we have the following useful identity 
\[
\e_{\m\n\r\s}= \frac{1}{g} g_{\m\a}g_{\n\b}g_{\r\g}g_{\s\l}
               \e^{\a\b\g\l}. 
\]
} of weight $+1$. 
On the other hand, the action (\ref{curv-bf}) possesses two kinds of 
invariance, given by
\eqa
\delta^{(1)}A^a_\mu \= -(D_\mu \theta)^a = 
- (\6_\mu \theta^a + f^{abc}A^b_\mu \theta^c) \ , \non
\delta^{(1)}B^a_{\mu\nu} \= f^{abc}\theta^b B^c_{\mu\nu} \ , 
\eqan{sym1}
and
\eqa
\delta^{(2)}A^a_\mu \= 0 \ , \non
\delta^{(2)}B^a_{\mu\nu} \= -(D_\mu \varphi_\nu - D_\nu \varphi_\mu)^a \ , 
\eqan{sym2}
$\theta^a$ and $\varphi^a_\mu$ are local parameters.
Now, by choosing a Landau gauge the gauge--fixing part of the action is 
given by \cite{zer3}  
\eqa
\S_{gf} \= - s \int_\MM d^4 x \sqrt{g} \bigg[g^{\mu\nu}\pa_\m \bar c^a 
A^a_\nu + g^{\mu\alpha}g^{\nu\beta}\pa_\a \bar\xi^a_\beta B^a_{\mu\nu}
+ g^{\mu\nu} \pa_\m \bar\phi^a \xi^a_\nu 
- g^{\mu\nu}\pa_\nu e^a \bar\xi^a_\mu 
- \bar\phi^a \lambda^a \bigg]  \non
&-& \frac{1}{2}\int_\MM d^4x ~\e^{\mu\nu\rho\sigma}f^{abc} \pa_\m \bar 
\x_\n^a \pa_\r \bar \x_\s^b \f^c , 
\eqan{gauge-fix} 
As already computed in \cite{zer3}, the extended BRS transformations of 
all fields introduced so far read
\eqa
sA^a_\mu \= -(D_\mu c)^a = -(\6_\mu c^a + f^{abc}A^b_\mu c^c) \ , \non
sB^a_{\mu\nu} \= -(D_\mu \xi_\nu - D_\nu \xi_\mu)^a 
+ f^{abc} c^b B^c_{\mu\nu} + \e_{\m\n\rho\sigma}f^{abc}
\sqrt{g}g^{\rho\a}g^{\sigma\b}(\6_\a\bar\x^b_\b)\f^c \ , \non
s\xi^a_\mu \= (D_\mu \phi)^a + f^{abc} c^b \xi^c_\mu  \ , \non
s\phi^a \= f^{abc} c^b \phi^c \ , \non
sc^a \= \half f^{abc} c^b c^c \ , \non
s\bar c^a \= b^a \ , \hspace{4.1cm} sb^a = 0 \ , \non
s\bar \x^a_\m \= h^a_\m \ , \hspace{4cm} sh^a_\m = 0 \ , \non
s\bar \f^a \= \omega^a \ , \hspace{4cm} s\omega^a = 0 \ , \non
s e^a \= \lambda^a \ , \hspace{4cm} s\lambda^a = 0 \ , \non 
s g_{\m\n} \= \hat g_{\m\n} \ , \hspace{3.8cm} s\hat g_{\m\n} = 0 \ . 
\eqan{brs-set}
The vector $\x^a_\m$ is the ghost field for the symmetry (\ref{sym2}) 
whereas $c^a$ is the ghost field for the gauge symmetry (\ref{sym1}).
$\f^a$ is the ghost for the ghost field $\x^a_\m$. Each of the
couples of fields
$(\bar c^a, b^a)$, $(\bar \x^a_\m, h^a_\m)$, $(\bar \f^a, \o^a)$ and
$(e^a, \l^a)$ contains an antighost and the corresponding Lagrange
multiplier fields.\\
We could extend the BRS transformations (see last line of (\ref{brs-set}))
by letting $s$ acting on the metric $g_{\m\n}$
because, at the level of the gauge fixed action $\S_{inv} + \S_{gf}$, 
the metric appears only in a BRS exact expression \cite{zer3}, a fact which
guarantee its non physical character. It turned out \cite{zer3} that
the BRS operator, constructed above, is nilpotent on--shell. more precisely,
\eq
s^2 B^a_{\m\n} = - \e_{\m\n\rho\sigma}f^{abc}
\frac{\d (\S_{inv}+\S_{gf})}{\d B^b_{\rho\sigma}}\phi^c~~~\hbox{and}~~~
s^2 = 0~~~\hbox{for the other fields} \ .
\eqn{inv1}

\begin{table}[h]
\renewcommand{\arraystretch}{1.2}
\begin{center}
\begin{tabular}{|c|c|c|c|c|c|c|c|c|c|c|c|c|c|c|c|c|}\hline
      &$A^a_\m$ &$B^a_{\m\n}$ &$c^a$ &$\bar c^a$ &$b^a$
      &$\x^a_\m$ &$\bar\x^a_\m$ &$h^a_\m$ &$\f^a$ &$\bar\f^a$
      &$\omega^a$ &$e^a$ &$\lambda^a$ &$g_{\m\n}$ &$\hat g_{\m\n}$
      \\ \hline
dim    &1 &2 &0 &2 &2 &1 &1 &1 &0 &2 &2 &2 &2 &0 &0 
\\ \hline
$\phi\pi$ &0 &0 &1 &-1 &0 &1 &-1 &0 &2 &-2 &-1 &0 &1 &0 &1 
\\ \hline
weight  &0 &0 &0 &0 &0 &0 &0 &0 &0 &0 &0 &0 &0 &0 &0 
\\ \hline
\end{tabular}
\\
\vspace{0.5cm}
\small{Table 1: Dimensions, ghost numbers and weights of the fields.}
\end{center}
\end{table}

\section{The on--shell analysis}

In the flat space--time limit the authors of \cite{sor2} constructed,
besides the BRS transformations, a further symmetry of the gauge 
fixed action, the so called vector supersymmetry--like transformations.
Their analysis was generalized \cite{zer3} to the case of a curved 
space--time admitting a covariantly constant vector field. In both cases 
the supersymmetry--like transformations were rigid transformations.
In this paper we make a further step and try to construct a {\it local}
supersymmetry--like transformations (at least on--shell), 
so let us begin by proposing the following transformations
\eqa
\d_{(\eta)}A^a_\mu \= -\e_{\m\n\rho\sigma}\eta^\n \sqrt g
g^{\rho\a}g^{\sigma\b}\6_\a\bar\xi^a_\b \ , \non
\d_{(\eta)}B^a_{\m\n} \= -\e_{\m\n\rho\sigma}\eta^\rho \sqrt g
g^{\sigma\a}\6_\a\bar c^a \ , \non
\d_{(\eta)}c^a \= -\eta^\mu A^a_\mu \ , \non
\d_{(\eta)}\bar c^a \= 0 \ , \non
\d_{(\eta)}b^a \= \LL_\eta\bar c^a \ , \non
\d_{(\eta)}\x^a_\m \= \eta^\n B^a_{\m\n} \ , \non
\d_{(\eta)}\bar\x^a_\m \= 0, \non
\d_{(\eta)}h^a_\m \= \LL_\eta\bar\x^a_\m , \non
\d_{(\eta)}\f^a \= \eta^\m \x^a_\m \ , \non
\d_{(\eta)}\bar \f^a \= 0 \ , \non
\d_{(\eta)}\omega^a \= \LL_\eta\bar \f^a \ , \non
\d_{(\eta)}e^a \= 0 \ , \non
\d_{(\eta)}\lambda^a \= \LL_\eta e^a \ , \non
\d_{(\eta)}g_{\m\n} \= 0 \ , \non
\d_{(\eta)}\hat g_{\m\n} \= \LL_\eta g_{\m\n} \ ,
\eqan{susy}
where $\LL_\eta$ is the Lie derivative and $\eta^\m$ is the vector 
parameter of the transformations with ghost number $+2$. 
It turns out that the on--shell algebra takes the following form
\eq
\lbrace s,\ \d_{(\eta)} \rbrace = \LL_\eta + ~~~\hbox{equ. of motion}
\eqn{aklsdj}
for all fields, except for the antisymmetric tensor field $B^a_{\m\n}$ 
where we get
\eq
\lac s , \d_{(\eta)} \rac B^a_{\m\n} = \LL_\eta B^a_{\m\n}
+\e_{\m\n\r\s} \eta^\r \frac{\d \S}{\d A^a_\s}
+\e_{\m\n\r\s} \eta^\r f^{abc} \sqrt g g^{\s\a} \pa_\a  
\bar\f^b \f^c \ .
\eqn{algebra-B}
Now to repair this shortcoming we add to the gauge fixed action the 
following expression
\eq
\S_{K,M}=\int d^4x (K_\m^a \Xi^{a\m}- M^a_\m s \Xi^{a\m} )
\eqn{sigmakm}
with
\eq 
\Xi^{a\m}=-f^{abc} \sqrt g g^{\m\a}\pa_\a \bar \f^b\f^c. 
\eqn{Xi}
The two auxiliary fields $K^a_\m$ and $M^a_\m$ transform under the
BRS and $\d_{(\eta)}$ operators according to 
\eq
\ba{llll}
s M^a_\m &=& K^a_\m,  ~~~&~~~   s K^a_\m = 0, \\
\d_{(\eta)} M^a_\m &=& 0,  ~~~&~~~ \d_{(\eta)} K^a_\m = \LL_\eta M^a_\m.
\ea
\eqn{brskm}
In an easy way we can prove that 
\eq
\lac s,\ \d_{(\eta)} \rac = \LL_\eta ~ ,
\eqn{kusdfgfi95hwj}
for the two auxiliary fields.
\begin{table}[h]
\renewcommand{\arraystretch}{1.2}
\begin{center}
\begin{tabular}{|c|c|c|}\hline
      &$K^a_\m$ &$M^a_\m$ \\ \hline
dim    &1 &1
\\ \hline
$\phi\pi$ &0 &-1
\\ \hline
weight  &0 &0 
\\ \hline
\end{tabular}
\\
\vspace{0.5cm}
\small{Table 2: Dimensions, ghost numbers and weights.}
\end{center}
\end{table}
The advantage of introducing the auxiliary fields $K^a_\m$ and $M^a_\m$ is
that (\ref{algebra-B}) will get the same form as (\ref{aklsdj}). Indeed,
\eq
\lac s , \d_{(\eta)} \rac B^a_{\m\n} = \LL_\tau B^a_{\m\n}
+\e_{\m\n\r\s} \eta^\r \bigg( \frac{\d \S}{\d A^a_\s} -
\frac{\d \S}{\d K^a_\s} \bigg)
\eqn{corralgeb}
So, in this way we have constructed an on--shell local 
supersymmetry--like transformations which anticommute with
the BRS operator and lead to Lie derivative (\ref{aklsdj}).
Next, we want to know if the transformations, 
generated by the operator $\d_{(\eta)}$, give rise to a symmetry of 
the gauge fixed action. This question is investigated in the next 
section where we also display the off--shell algebra.

\section{The off--shell analysis}

In order to generalize the above results to the off--shell level
we first introduce external sources which couple to the non--linear
BRS transformations (\ref{brs-set})
\eqa
\S_{ext} \= \int_\MM d^4x \Big[\Omega^{a\m}(sA^a_\m) + 
\gamma^{a\m\n}(sB^a_{\m\n})
+L^a(sc^a)+D^a(s\f^a)+\rho^{a\m}(s\x^a_\m)\Big] \non
&+&\in \Big[\frac{1}{2}\e_{\m\n\rho\sigma}f^{abc}\gamma^{a\m\n}
\gamma^{b\rho\sigma}\f^c \Big] \ ,
\eqan{ext-action}
with dimensions, weights and ghost numbers as given in table 3.
\begin{table}[h]
\renewcommand{\arraystretch}{1.2}
\begin{center}
\begin{tabular}{|c|c|c|c|c|c|}\hline
      &$\gamma^{a\m\n}$ &$\Omega^{a\m}$ &$L^a$ &$D^a$ &$\rho^{a\m}$
      \\ \hline
dim    &2 &3 &4 &4 &3 
\\ \hline
$\phi\pi$ &-1 &-1 &-2 &-3 &-2 
\\ \hline
weight  &1 &1 &1 &1 &1 
\\ \hline
\end{tabular}
\\
\vspace{0.5cm}
\small{Table 3: Dimensions, ghost numbers and weights of the 
external sources.}
\end{center}
\end{table} 

\noindent
From the transformations \equ{susy} and the second line of 
(\ref{brskm}) (and due to presence of the
external sources) we get the Ward operator $\bar \WW^S_{(\eta)}$ 
such that
\eqa
\bar \WW^S_{(\eta)} \= \in \Bigg[ 
-\e_{\m\n\rho\sigma}\eta^\n (\g^{a\r\s} + \sqrt g   
g^{\rho\a}g^{\sigma\b}\6_\a\bar\xi^a_\b) \dd{A^a_\mu} 
-\eta^\mu A^a_\mu \dd{c^a} + \LL_\eta\bar c^a \dd{b^a} - \non
&-&\e_{\m\n\rho\sigma}\eta^\rho (\O^{a\s} +\sqrt g
g^{\sigma\a}\6_\a\bar c^a ) \dd{B^a_{\m\n}}    
+\eta^\n B^a_{\m\n} \dd{\x^a_\m}   
+\LL_\eta\bar\x^a_\m \dd{h^a_\m} 
+\eta^\m \x^a_\m \dd{\f^a} + \non 
&+& \LL_\eta\bar \f^a\dd{\omega^a} 
+\LL_\eta e^a \dd{\lambda^a} 
+ \LL_\eta g_{\m\n} \dd{\hat g_{\m\n}} 
-\eta^\mu D^a \dd{\rho^{a\m}}
-\eta^\mu L^a \dd{\Omega^{a\m}} - \non
&-& \eta^\mu \rho^{a\n} \dd{\gamma^{a\m\n}} +
 (\LL_\eta M^a_\m - \e_{\m\n\r\s} \eta^\n \g^{a\r\s})\dd{K^a_\m}
\Bigg] \ .
\eqan{ward-operator-susy}
After tedious calculations, the corresponding Ward identity takes the form
\eq
\bar\WW^S_{(\eta)} (\S_{inv}+\S_{gf}+\S_{K,M}+\S_{ext}) = 
\Delta^{cl}_{(\eta)} \ ,
\eqn{ward-susy}
where the classical breaking $\D^{cl}_{(\eta)}$ split in linear and 
quadratic parts 
\eq
\Delta^{cl}_{(\eta)} = \D^L_{(\eta)} + \D^Q_{(\eta)}.
\eqn{classical-breaking}
with,
\eqa
\Delta^L_{(\eta)} \= \in \Big[ -\gamma^{a\m\n}\LL_\eta B^a_{\m\n}
-\Omega^{a\m}\LL_\eta A^a_{\m}+L^a\LL_\eta c^a-D^a\LL_\eta \f^a
+\rho^{a\m}\LL_\eta \x^a_{\m} - \non
&-&\e_{\m\n\rho\sigma}\Omega^{a\m} \eta^\n 
s(\sqrt g g^{\rho\a}g^{\sigma\b}\6_\a\bar\x^a_\b)
- \e_{\m\n\rho\sigma}\gamma^{a\m\n} \eta^\rho 
s(\sqrt g g^{\sigma\a}\6_\a\bar c^a)
\Big]
\eqan{linear-breaking}
and
\eqa
\D^Q_{(\eta)} &=& \int d^4 x \Big[
- \e^{\m\n\r\s} s (g_{\m\a}\eta^\a \pa_\n \bar c^a \pa_\r \bar \x^a_\s)
- s (\sqrt g g^{\m\n} \pa_\m \bar \f^a \eta^\a B^a_{\a\n})
- s (\sqrt g \bar \f^a \LL_\eta e^a) - \non
&-& s (f^{abc} \sqrt g g^{\m\a} M^a_\m \pa_\a \bar \f^b \eta^\n
\x^c_\n) \Big] \ .
\eqan{quadratic-breaking}
The nonlinear breaking (\ref{quadratic-breaking}) is quadratic
in the quantum fields, then it is not harmless in the context of 
the renormalization procedure.\\
To eliminate the nonlinear expression in (\ref{classical-breaking}) we
first add to the action the BRS exact integral
\eqa
\S_1 &=& \int d^4 x \Big( -L_\m \Y^\m + R_\m s \Y^\m 
                           +W^\m \Ps_\m - Z^\m s \Ps_\m 
                       + I^a_\m \T^{a\m} -  \non
                      &-& J^a_\m s \T^{a\m} 
                       + P^a \LA^a + Q^a s \LA^a \Big)
\eqan{lrlrlr}
such that
\eq
\ba{rcl}
\Y^\m &=& \e^{\m\n\r\s} \pa_\n \bar c^a \pa_\r \bar \x^a_\s , \\
\Ps_\m &=& \bar \f^a \pa_\m e^a , \\
\T^{a\m} &=& \sqrt g g^{\m\r} \pa_\r \bar \f^a , \\
\LA^a &=& f^{abc} \sqrt g g^{\m\r} M^b_\m \pa_\r \bar \f^c .
\ea
\eqn{jkhfglu}
All new auxiliary fields introduced in (\ref{lrlrlr}) transform in
a BRS doublets 
\eq
\ba{llll}
s R_\m &=& L_\m ,  ~~~~&~~~~ s L_\m =0, \\
s Z^\m &=& W^\m ,  ~~~~&~~~~ s W^\m =0, \\
s J^a_\m &=& I^a_\m ,  ~~~~&~~~~ s I^a_\m =0, \\
s Q^a &=& P^a ,  ~~~~&~~~~ s P^a =0.
\ea
\eqn{ha98asd}
Furthermore, under the $\d_{(\eta)}$ operation, they transform as
\eq
\ba{llll}
\d_{(\eta)} R_\m &=& g_{\m\r} \eta^\r ,  ~~~~&~~~~
\d_{(\eta)} L_\m = \LL_\eta R_\m - \hat g_{\m\r} \eta^\r , \\
\d_{(\eta)} Z^\m &=& - \sqrt g \eta^\m ,  ~~~~&~~~~
\d_{(\eta)} W^\m = \LL_\eta Z^\m + \eta^\m s \sqrt g , \\
\d_{(\eta)} J^a_\m &=& \eta^\n B_{\m\n}^a ,  ~~~~&~~~~
\d_{(\eta)} I^a_\m = \LL_\eta J^a_\m - s (\eta^\n B_{\m\n}^a), \\
\d_{(\eta)} Q^a &=& \eta^\m \x^a_\m ,  ~~~~&~~~~
\d_{(\eta)} P^a = \LL_\eta Q^a - s (\eta^\m \x^a_\m).
\ea
\eqn{834oihuer}

\begin{table}[h]
\renewcommand{\arraystretch}{1.2}
\begin{center}
\begin{tabular}{|c|c|c|c|c|c|c|c|c|}\hline
      & $L_\m$ & $R_\m$ & $W^\m$ & $Z_\m$ & $I^a_\m$ & $J^a_\m$ &
      $P^a$ & $Q^a$ 
      \\ \hline
dim    &-1 &-1 &-1 &-1 &1 &1 &0 &0  
\\ \hline
$\phi\pi$ &2 &1 &2 &1 &2 &1 &3 &2  
\\ \hline
weight  &0 &0 &1 &1 &0 &0 &0 &0  
\\ \hline
\end{tabular}
\\
\vspace{0.5cm}
\small{Table 1: Dimensions, ghost numbers and weights of auxiliary 
fields.}
\end{center}
\end{table}

\noindent
For all auxiliary fields, the anticommutator of the BRS operator $s$
and $\d_{(\eta)}$ closes on the Lie derivative plus equations of motion.\\
On the other hand, the Ward identity (\ref{ward-susy}) gets promoted to
\eq
\WW^S_{(\eta)}(\S) = \bar \WW^S_{(\eta)}(\S) +  \VV^S_{(\eta)}(\S)
= \D^L_{(\eta)},
\eqn{ljkafg}
where $\VV^S_{(\eta)}(\S)$ is nonlinear, and its expression is
\eqa
\VV^S_{(\eta)}(\S) &=& \int d^4x \Bigg(
g_{\m\r}\eta^\r \ds{R_\m}
+ (\LL_\eta R_\m - \hat g_{\m\r} \eta^\r) \ds{L_\m}
- \sqrt g \eta^\m \ds{Z^\m}
+ \eta^\n B^a_{\m\n} \ds{J^a_\m} + \non
&+& (\LL_\eta Z^\m + \eta^\m s \sqrt g) \ds{W^\m}              
+ (\LL_\eta J^a_\m - \eta^\n \ds{\g^{a\m\n}})\ds{I^a_\m}
+ \eta^\m \x^a_\m \ds{Q^a} + \non
&+& (\LL_\eta Q^a - \eta^\m \ds{\r^{a\m}} )\ds{P^a}  
\Bigg) 
\eqan{jklerliertkljdfgnmf}
The complete gauge fixed action is now given by
\eq
\Sigma=\S_{inv}+\S_{gf}+\S_{K,M}+\S_{ext} + \S_1 . 
\eqn{action-classical}
It obeys the Slavnov identity
\eq
\SS(\Sigma) = 0 \ ,
\eqn{slavnov}
where 
\eqa
\SS(\Sigma)\= \in \Bigg(\ds{\gamma^{a\m\n}}\ds{B^a_{\m\n}}
+\ds{\Omega^{a\m}}\ds{A^a_{\m}}+\ds{L^a}\ds{c^a}  
+\ds{D^a}\ds{\f^a}+\ds{\rho^{a\m}}\ds{\x^a_{\m}}+ \non
&+&b^a \ds{\bar c^a}+h^a_\m \ds{\bar\x^a_\m}
+\omega^a \ds{\bar \f^a}+\lambda^a \ds{e^a}
+ \hat g_{\m\n} \ds{g_{\m\n}} 
+ K_\m^a \ds{M_\m^a}+ L_\m \ds{R_\m} + \non
&+& W^\m \ds{Z^\m} + I^a_\m \ds{J^a_\m}
+ P^a \ds{Q^a}
\Bigg) \ .
\eqan{slavnov-identity}
It is straightforward to verify that the corresponding linearized 
Slavnov operator is given by
\eqa
\SS_\Sigma \= \in \Bigg(\ds{\gamma^{a\m\n}}\dd{B^a_{\m\n}}
+\ds{B^a_{\m\n}}\dd{\gamma^{a\m\n}}
+\ds{\Omega^{a\m}}\dd{A^a_{\m}}+\ds{A^a_{\m}}\dd{\Omega^{a\m}}+ \non
&+&\ds{L^a}\dd{c^a}+\ds{c^a}\dd{L^a}  
+\ds{D^a}\dd{\f^a}+\ds{\f^a}\dd{D^a}
+\ds{\rho^{a\m}}\dd{\x^a_{\m}}+\ds{\x^a_{\m}}\dd{\rho^{a\m}}+ \non
&+&b^a \dd{\bar c^a}+h^a_\m \dd{\bar\x^a_\m}
+\omega^a \dd{\bar \f^a}+\lambda^a \dd{e^a}
+ \hat g_{\m\n} \dd{g_{\m\n}} + K^a_\m \frac{\d}{\d M^a_\m} 
+ L_\m \frac{\d}{\d R_\m} + \non
&+& W^\m \frac{\d}{\d Z^\m} + I^a_\m 
\frac{\d}{\d J^a_\m} + P^a \frac{\d}{\d Q^a} 
\Bigg)
\eqan{linear-slavnov-operator}
At the functional level, the invariance of the classical action
(\ref{action-classical})
under diffeomorphisms can be expressed by an unbroken Ward identity
\eq
\WW^D_{(\e)} \Sigma = 0 \ ,
\eqn{ward-diff}
where $\WW^D_{(\e)}$ denotes the corresponding Ward operator,
\eq
\WW^D_{(\e)} = \in \sum_{f} \big( 
\LL_\e f \big) \dd{f} \ ,
\eqn{ward-operator-diff}
for all fields $f$. The vector parameter of the 
diffeomorphism transformations is denoted by $\e^\m$, and it carries
ghost number $+1$.\\
Next, we display the complete nonlinear algebra 
of the Slavnov operator and the Ward operator $\WW^D_{(\e)}$.
To this end, let $\Gamma$ be an arbitrary functional depending on 
the fields of the model, then
\eqa
\SS_\Gamma \SS(\Gamma) \= 0 \ , \non
\SS_\Gamma \WW^D_{(\e)} \Gamma + \WW^D_{(\e)}\SS(\Gamma) \= 0 \ , \non
\lac \WW^D_{(\e)} , \WW^D_{(\e')} \rac \Gamma \= 
-\WW^D_{(\lac\e,\e'\rac)} \Gamma \ .
\eqan{nonlinear-algebra1}
Now if the functional $\Gamma$ is a solution of the Slavnov identity
and of the Ward identity of diffeomorphisms, then 
the off--shell algebra \equ{nonlinear-algebra1} reduces to the linear 
algebra
\eqa
\SS_\Sigma\SS_\Sigma  \= 0 \ , \non
\lac \SS_\Sigma , \WW^D_{(\e)} \rac \= 0 \ , \non
\lac \WW^D_{(\e)} , \WW^D_{(\e')} \rac \= -\WW^D_{(\lac\e,\e'\rac)} \ , \non
\eqan{off-shell-algebra}
with the Lie brackets 
\eq
 \{\e,\e'\}^\m = \LL_\e\e'^\m 
\eqn{lie-def}
Contrary to other topological field theories \cite{zer2} \cite{zer1}, the 
anticommutator of the linearized Slavnov operator 
(\ref{linear-slavnov-operator})
with the linearized\footnote{First one has to linearize $\WW^S_{(\eta)}$
and then let the anticommutator $\lbrace \SS_\S, \WW^S_{(\eta)} 
\rbrace$ acting on the different fields.}
$\WW^S_{(\eta)}$ Ward operator does not give rise to the diffeomorphisms
Ward operator. Indeed, for certain fields we get    
$\lbrace \SS_\S, \WW^S_{(\eta)} \rbrace = \WW^D_{(\e)}$ and for other
fields
the anticommutator does not close on $\WW^D_{(\e)}$. As a simple example,
one can let the above anticommutator acting on the two auxiliary fields
$M^a_\m$ and $K^a_\m$.\\
The main result of this paper is that we could, after introducing 
auxiliary fields, construct a local and nonlinear symmetry of the
$4D$ antisymmetric tensor field model in a curved manifold. 
The next natural step to do is to show the renormalizability of the
new symmetry. It turns out that if this symmetry is valid at the
quantum level then it would be 
very useful in showing the finiteness of the model in 
a big class of curved manifolds. This would be the generalisation
of the results of \cite{zer3}.

\section*{Acknowledgment} 

I would like to thank the ``Fonds zur F\"orderung der Wissenschaftlichen 
Forschung'' for the financial support: contract grant number P11582-PHY.


\end{document}